\documentclass[lettersize,journal]{IEEEtran}
\usepackage{amsmath,amsfonts}
\usepackage{algorithmic}
\usepackage{algorithm}
\usepackage{array}
\usepackage[caption=false,font=normalsize,labelfont=sf,textfont=sf]{subfig}
\usepackage{textcomp}
\usepackage{stfloats}
\usepackage{url}
\usepackage{verbatim}
\usepackage{graphicx}
\usepackage{cite}
\usepackage{orcidlink}

\begin{document}

\title{Evaluating the Deformation Measurement Accuracy Using Low-SNR Radars for Future InSAR Missions}

\author{Emre Havazli\orcidlink{0000-0002-1236-7067}, Shadi Oveisgharan, Michael Denbina, Brian Hawkins 
\thanks{E. Havazli, S. Oveisgharan, M. Denbina and B. Hawkins are with Jet Propulsion Laboratory,
California Institute of Technology, Pasadena, CA, USA. (email: emre.havazli@jpl.nasa.gov)}

}

\IEEEpubid{\copyright~2025. California Institute of Technology. Government sponsorship acknowledged. }

\maketitle
\begin{abstract}
Interferometric Synthetic Aperture Radar (InSAR) is a powerful tool for monitoring surface deformation with high precision. However, low Signal-to-Noise Ratio (SNR) conditions, common in regions with low backscatter, can degrade phase coherence and compromise displacement accuracy. In this study, we quantify the impact of low-SNR conditions on InSAR-derived displacement using L-band UAVSAR data collected over the San Andreas Fault and Greenland ice sheet. We simulate low-SNR conditions by degrading the Noise-Equivalent Sigma Zero (NESZ) to $-15~\mathrm{dB}$ and assess the resulting effects on interferometric coherence, phase unwrapping, and time series inversion. The displacement accuracy of 4mm in single interferogram can be achieved by taking looks for the signal decorrelation of 0.6 and SNR between -9dB to -10dB. Our findings indicate that even under low-SNR conditions, a velocity precision of $0.5~\mathrm{cm/yr}$ can be achieved in comparison to high-SNR conditions. By applying multilooking with an 8x8 window, we significantly improve coherence and eliminate this bias, demonstrating that low-SNR systems can achieve comparable precision to high-SNR systems at the expense of spatial resolution. These results have important implications for the design of future cost-effective SAR missions, such as Surface Deformation and Change (SDC), and the optimization of InSAR processing techniques in challenging environments.
\end{abstract}


\section{Introduction}
\IEEEPARstart{I}{nterferometric} Synthetic Aperture Radar (InSAR) techniques are widely used for high-precision monitoring of surface deformation, enabling detailed mapping of geophysical processes such as tectonic movements (e.g., \cite{wright2004insar, bekaert2016network}), glacier dynamics (e.g., \cite{leinss2021tandem, feng2023improving}), and subsidence (e.g., \cite{chaussard2021over, pacheco2015application}). InSAR measures phase differences between radar acquisitions to detect surface displacements over time (\cite{massonnet1998radar, hanssen2001radar}).

One of the primary challenges in InSAR is signal decorrelation, which leads to a loss of phase information and reduced measurement accuracy. Decorrelation arises from different sources including temporal changes on the surface, geometric misalignments, and low Signal-to-Noise Ratio (SNR). In this study, we use the term Low SNR for the backscattered power close to, or lower than the system's Noise-Equivalent Sigma Zero (NESZ). Low SNR is observed in regions with low reflectivity such as snow, ice, or arid surfaces \cite{zebker1992decorrelation}.

Low SNR conditions increase phase noise in interferograms, complicating phase unwrapping and introducing errors in displacement estimates \cite{goldstein1988satellite}. While the impact of low SNR on phase stability is recognized, comprehensive quantification using real-world datasets remains limited. Previous studies have focused on theoretical models or controlled simulations, leaving gaps in understanding the practical implications of low-SNR conditions \cite{rocca2007modeling, lu2007insar}.

This study addresses that gap by evaluating the impact of low SNR on displacement accuracy using L-band UAVSAR data. UAVSAR's longer wavelength (24 cm) helps maintain coherence in challenging environments like vegetated and snow-covered regions \cite{fore2015uavsar}. However, even L-band data are susceptible to noise at low SNR levels. We simulate low-SNR conditions by degrading the NESZ of UAVSAR data from its nominal $-50\mathrm{dB}$ to $-15~\mathrm{dB}$ \cite{hawkins2018application}, allowing for a controlled assessment of phase stability and displacement accuracy under degraded conditions.

Our analysis focuses on UAVSAR data from two contrasting environments: the tectonically active San Andreas Fault and the rapidly deforming Greenland ice sheet. By comparing the original and noise-degraded datasets, we quantify the effects of low SNR on single interferogram phase accuracy, phase unwrapping, and time series inversion. The results offer insights into optimizing InSAR processing for low-backscatter regions and designing cost-effective future SAR missions such as NASA's Surface Deformation and Change (SDC) project.

\section{Data and Methodology}
\subsection{UAVSAR Data and Simulated Low NESZ Data}\label{sec-method}\IEEEpubidadjcol
We use the data collected by the UAVSAR instrument for our analysis. UAVSAR operates in the L-band frequency, which has a wavelength of approximately 24 cm. This longer wavelength is particularly advantageous for preserving temporal coherence compared with shorter-wavelength bands such as C-band or X-band, especially in environments with significant vegetation (\cite{rosen2007uavsar, hensley2008uavsar, bekaert2019exploiting}). By leveraging UAVSAR datasets, we aim to assess the impact of low SNR on deformation estimations.

For our study, we selected two regions, both with backscattered power levels below -15 dB, allowing us to evaluate the performance of the InSAR products with a -15 dB NESZ radar. The first site is a single interferogram over Greenland whereas the second site is over San Andreas Fault with a time series stack. The two study sites and available UAVSAR data for each site are explained below in \ref{sec-greenland} and \ref{sec-SanAnd}.

The first step in our analysis is to obtain both the ``original'' and ``noisy'' SLCs. The original SLCs are UAVSAR acquisitions available publicly, while we generated the ``noisy'' SLCs  by adjusting the NESZ of original data to a constant -15 dB. To achieve this, we reference the NESZ vs. incidence angle relationships from Fore et al. \cite{fore2015uavsar}, accounting for temporal variations in hardware and radiometric calibration. The NESZ curve is adjusted based on each UAVSAR flight line’s acquisition date and interpolated to the corresponding incidence angle for each pixel.

With the original NESZ determined, we compute the additional noise required to set the noisy stack’s NESZ to -15 dB. Unlike UAVSAR’s spatially varying NESZ, our synthetic noise is constant across the swath for simplicity. The real and imaginary noise components are independently sampled from Gaussian distributions with zero mean and standard deviation equal to:

\begin{equation}\label{eq1}
  \sqrt{N_{\text{noise}} - N_{\text{original}}}
\end{equation}

where \(N_{\text{noise}}\) and \(N_{\text{original}}\) (converted to linear units) represent the noise-added and original NESZ, respectively. The square root accounts for the conversion from power to amplitude. Independent random seeds ensure uncorrelated noise across SLCs.

\begin{figure}[hbp!]
    \centering
    \includegraphics[width=\linewidth]{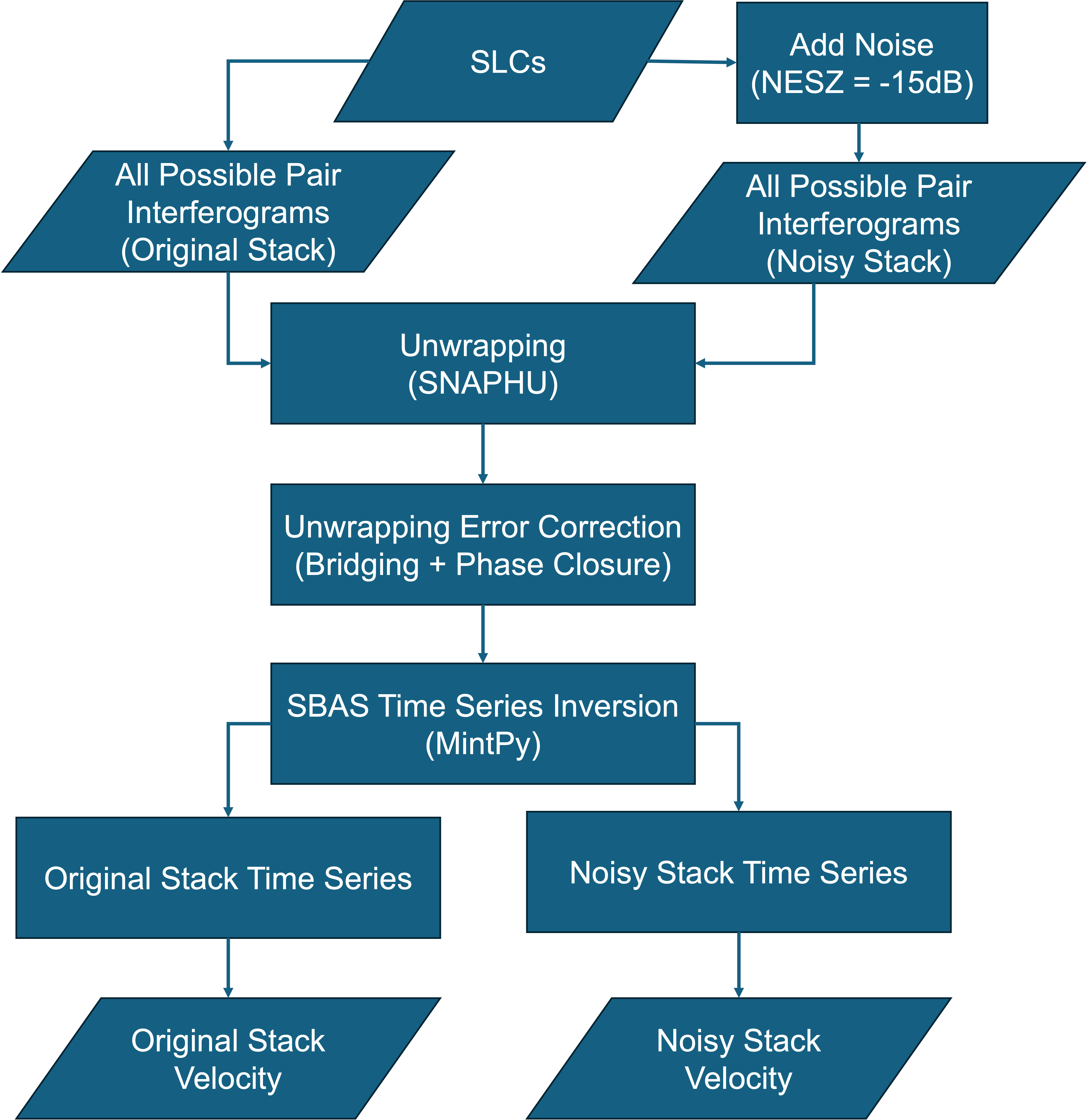}
    \caption{Workflow for processing UAVSAR SLCs, including noise simulation, interferogram generation, phase unwrapping, and time series inversion.}
    \label{fig-workflow}
\end{figure}

After obtaining both ``original'' and ``noisy'' time series SLC stacks over San Andreas Fault, we compute all possible pairwise interferograms independently for each stack. Both the original and noisy stacks undergo a rigorous processing workflow (Fig.\ref{fig-workflow}). Phase unwrapping is performed using SNAPHU \cite{chen2000network, chen2001two, chen2002phase}, employing methods optimized for minimizing errors in regions with complex deformation. This is followed by a bridging process to resolve phase discontinuities across connected components and phase closure corrections to mitigate cumulative unwrapping errors that could distort long-term displacement trends \cite{yunjun2019small}. These steps are critical for ensuring the reliability of the unwrapped phase, particularly in low-SNR conditions.

Next, we apply a Small BAseline Subset (SBAS) time series inversion using the MintPy software package \cite{yunjun2019small}. The SBAS approach refines surface deformation time series by leveraging a dense interferogram network, effectively reducing decorrelation effects. By comparing results from the original and noisy stacks, we assess the impact of reduced SNR on displacement accuracy, providing insights into the limitations and potential improvements of InSAR monitoring in low-backscatter regions.

\subsection{Case Study over the Greenland Ice Sheet} \label{sec-greenland}

\begin{figure}[hbp!]
    \centering
    \includegraphics[width=\linewidth]{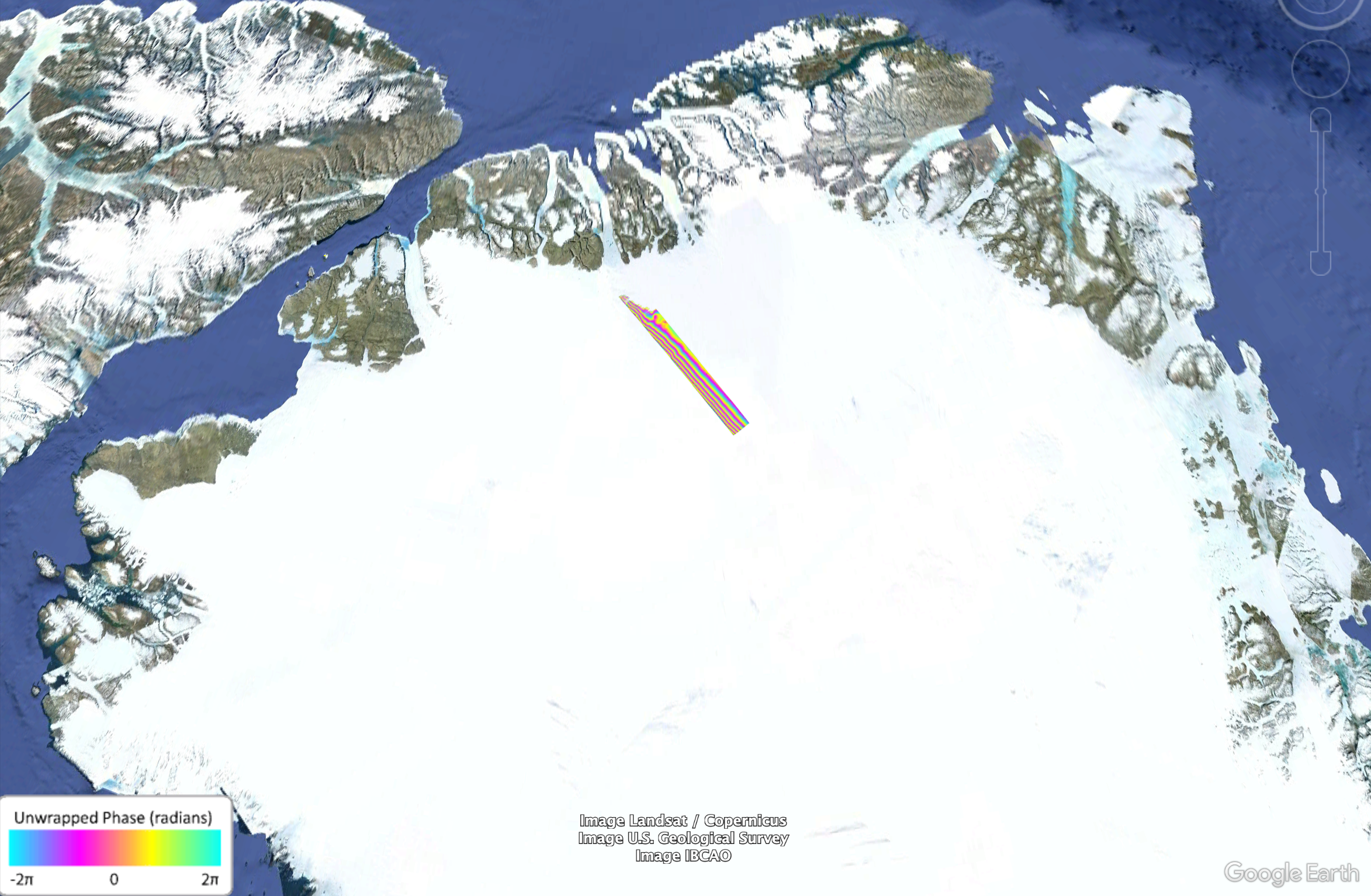}
    \caption{UAVSAR flight line Grnlnd 13801 over Northern Greenland}
    \label{grnland_fig}
\end{figure}

The first region selected for this study is the Northern Greenland Ice Sheet (shown in figure \ref{grnland_fig}), which is characterized by a rapid fringe rate in a single interferogram due to significant ice dynamics. This region was specifically chosen to evaluate the performance of the InSAR system in environments where substantial deformation occurs within a single interferogram, offering insights into the system’s ability to resolve fine-scale motion under low-SNR conditions. We used the UAVSAR flight line Grnlnd$\_$13801 over Greenland, with data acquired on 05/21/2009 and 05/29/2009 (Fig. \ref{Grnlnd_sigma_cor_int_unwdif_fig}).

\subsection{Case Study over the San Andreas Fault Zone}\label{sec-SanAnd}
The second region selected for this study is the San Andreas Fault zone, which is studied extensively using InSAR time series methods (e.g., \cite{ryder2008spatial, lundgren2009southern, aissa2024united}). This area is notable for its tectonic activity, characterized by slow and steady surface deformations that accumulate over time, making it an ideal candidate for long-term analysis. The San Andreas Fault region not only offers InSAR time series data with low and high backscattered power but also encompasses a diverse range of correlation values unlike Greenland Ice sheet. This range allows for a comprehensive assessment of the InSAR instrument's performance across varying signal conditions. The diverse range of correlation is due to wide range of time intervals between data acquisitions from couple of months to couple of years.

\begin{figure}[hbp!]
    \centering
    \includegraphics[width=\linewidth]{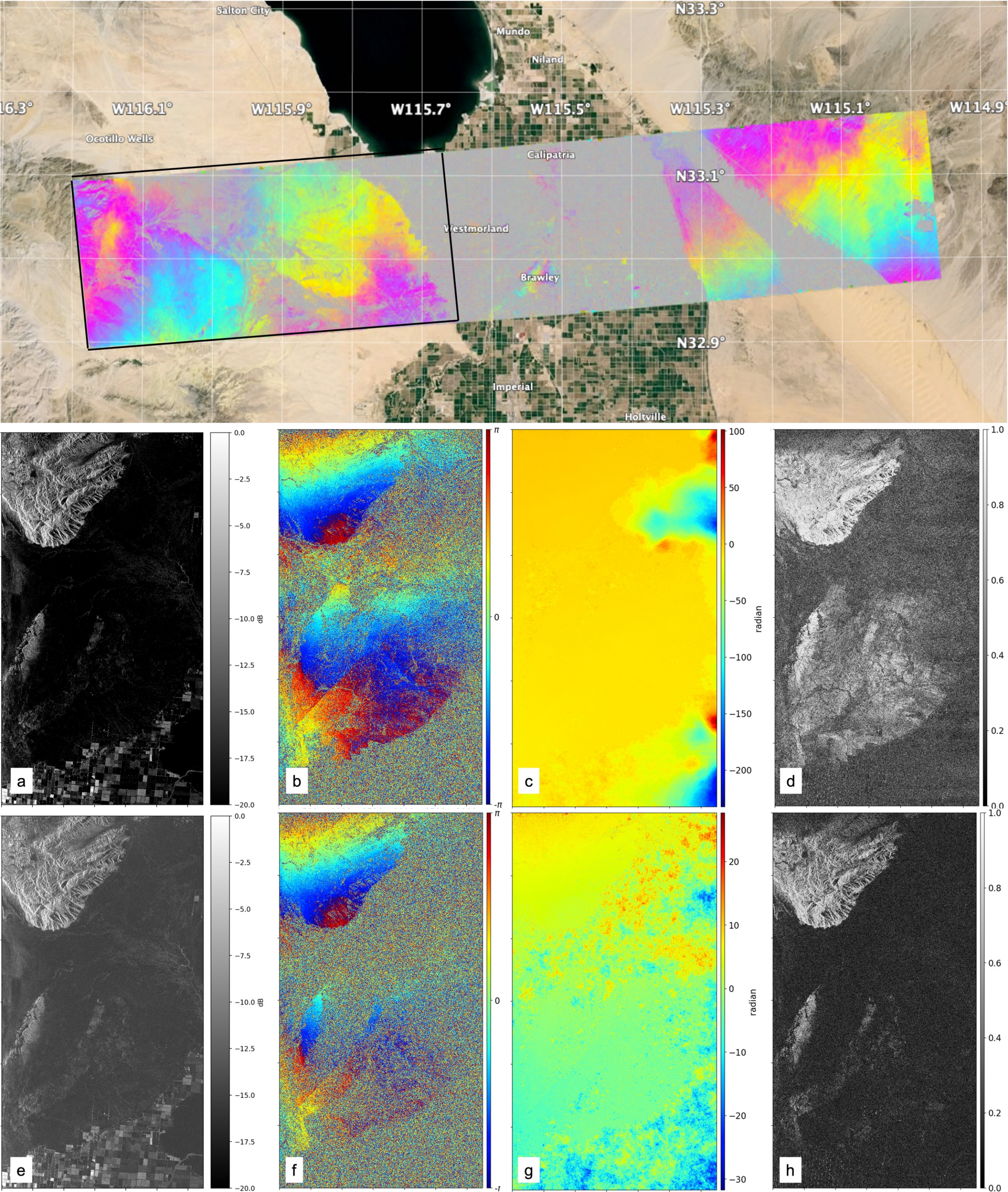}
    \caption{San Andreas Fualt UAVSAR: The top image shows a sample interferogram over our study area. The black box over the top image shows the subset area we used in our analysis. a, b, c, and d are the power, wrapped interferogram, unwrapped interferogram and coherence from a sample in the original stack. e, f, g, and h are the power, wrapped interferogram, unwrapped interferogram and correlation from the same sample in the noisy stack.}
    \label{fig-samples}
\end{figure}

For this study, we utilized UAVSAR data from deployment SanAnd$\_$08508, which covers the San Andreas Fault Zone (SAFZ) and the Salton Trough (Fig.~\ref{fig-samples}). Fig.~\ref{fig-samples} illustrates the geographic extent of the study area along with the original and noise-added observations for power, wrapped and unwrapped interferograms, and interferometric coherence for a single interferometric pair. The top image denotes the extent of our study area subset with a black frame. A noticeable degradation in power is evident when comparing the original (Fig.~\ref{fig-samples}a) and noise-added (Fig.~\ref{fig-samples}e) images. A pronounced loss of coherence is observed near the center of the scene, as seen in the wrapped interferograms (Figs.~\ref{fig-samples}b and \ref{fig-samples}f), which propagates as unwrapping errors in the corresponding unwrapped interferograms (Figs.~\ref{fig-samples}c and \ref{fig-samples}g). The correlation maps (Figs.~\ref{fig-samples}d and \ref{fig-samples}h) confirm widespread decorrelation across the interferogram after adding noise.


Our original data set consists of 14 acquisitions collected between April 2009 and November 2021. We started our analysis by generating the "noisy" data set by bringing the NESZ of the original acquisitions to -15dB as described in Section \ref{sec-method}. After obtaining both "original" and "noisy" data sets, we carried out interferogram generation by making all possible pair interferograms for both original and noisy data sets individually. After the interferogram generation, we continue with phase unwrapping and time series inversion and generate velocity fields from both original and noisy data sets.

\section{Results} \label{sec-results}
In this section, we evaluate the performance of a system with low NESZ. In Section \ref{sec-results_single_int}, we assess the performance of a single interferogram using both "original" and "noisy" data over Greenland and the San Andreas Fault. In Section \ref{sec-results_stack_int}, we explore how time series data can help compensate for some of the lost information in a system with low NESZ.

\subsection{Performance of Single Interferogram for a low NESZ Radar} \label{sec-results_single_int}

In this section, we compare the "noisy" and "original" interferometric phases to evaluate the phase error in low NESZ systems. The first single interferogram we used is over Greenland.

\begin{figure}[!hbp]
\begin{center}
\includegraphics[width=\linewidth]{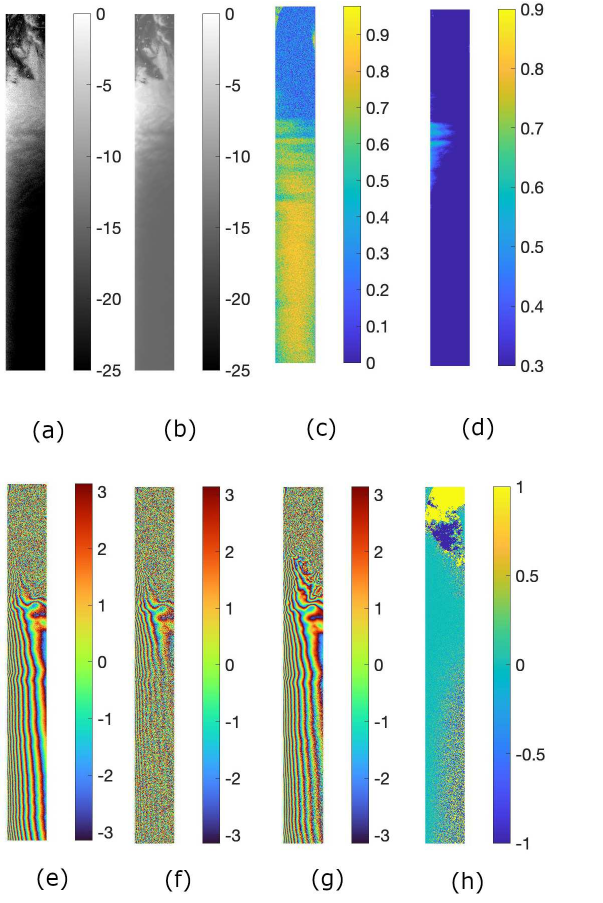}
\caption{Greenland UAVSAR (a) backscattered power, resolution: 7m, (b) simulated low NESZ backscatetred power, resolution: 56m, (c) correlation, resolution: 7m, (d) simulated low NESZ correlation, resolution: 56m, (e) interferogram, resolution: 7m, (f) simulated low NESZ interferogram, resolution: 7m, (g) simulated low NESZ interferogram, resolution: 56m, (h) difference between original and simulated low NESZ unwrapped phase, resolution: 56m  } \label{Grnlnd_sigma_cor_int_unwdif_fig}
\end{center}
\end{figure}

Figures \ref{Grnlnd_sigma_cor_int_unwdif_fig}(a), (c), and (e) show the backscattered power, intereferometric correlation, and interferometric phase, respectively. As observed in these figures, the backscattered power is very low, and the correlation is very high in the lower half of the images. We generated the "noisy" data as described in Section \ref{sec-method}. The corresponding "noisy" backscattered power, correlation, and interferometric phase are shown in Figures \ref{Grnlnd_sigma_cor_int_unwdif_fig}(b), (d), and (f), respectively. As seen in these figures, the "noisy" system significantly reduces the correlation in the lower half of the image. Additionally, the dense fringes in the lower half of the "original" interferogram become blurred in the "noisy" interferogram. The difference between the "noisy" and "original" unwrapped phases in the dark regions is biased due to poor unwrapping. By applying multilooking with an 8x8 window, we can reduce speckle noise and consequently improve the SNR and correlation. Figure \ref{Grnlnd_sigma_cor_int_unwdif_fig}(g) shows the "noisy" interferogram with 8x8 looks (56 m resolution). As seen in this image, the lost fringes are now visible again. The difference between the "noisy" and "original" unwrapped phases is shown in Figure \ref{Grnlnd_sigma_cor_int_unwdif_fig}(h). We will quantify this error further in the remainder of this section.

\begin{figure}[!hbp]
\begin{center}
\includegraphics[width=\linewidth]{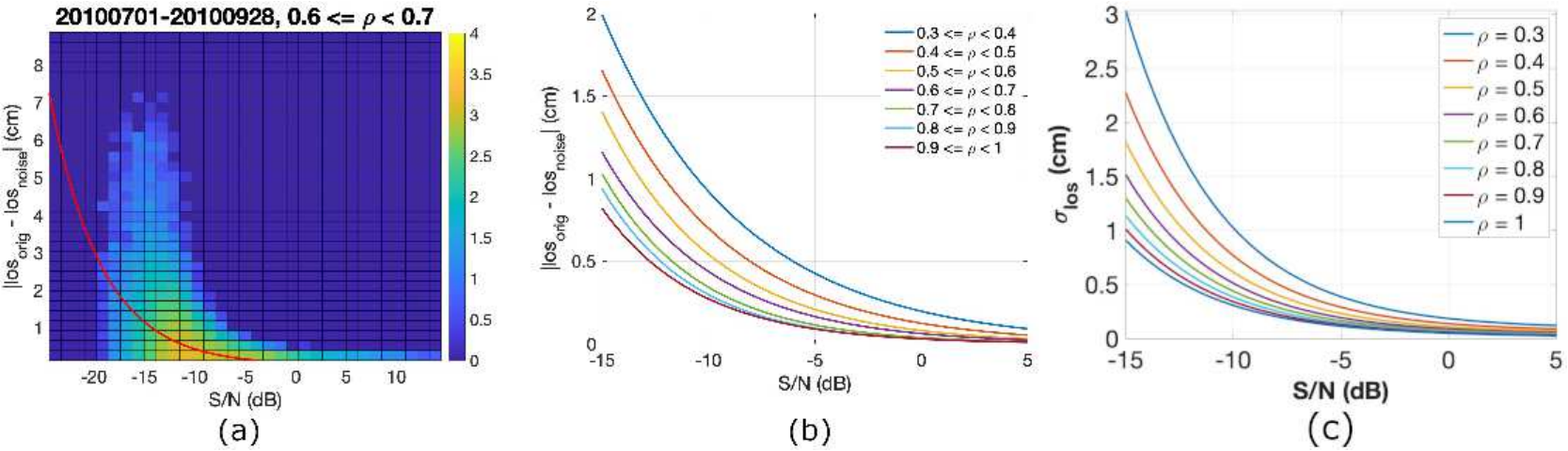}
\caption{(a) 2D histogram of Line of Sight deformation (between 07/01/2010-09/28/2010) difference between "original" UAVSAR and "noisy" data versus SNR for $0.6 <= \rho <0.7$ over the San Andreas Fault. The red curve shows the exponential fit to the 2D LOS difference. (b) The exponential fit curve for $\rho$. (c) Theoretical standard deviation of LOS error vs S/N (dB). Different colors show the "original" image correlation. } \label{err_snr_hist6_sanand1_fig}
\end{center}
\end{figure}

As mentioned in Section \ref{sec-SanAnd}, we also have a stack of interferograms over the San Andreas Fault. Unlike Greenland, the stack of interferograms in this region shows a wide range of (backscattered power, correlation) values (Fig. \ref{fig-samples}). As the time between single interferograms increases from a few months to 12 years, coherence decreases, especially in regions with low backscattered power. As a result, we observe a diverse range of coherence values in the low backscatter regions of the San Andreas Fault. In this section, we use four single interferograms over the San Andreas Fault to cover a broad range of coherence. The time intervals for these four interferograms are 3 months (07/01/2010-09/28/2010), 10 months (07/01/2010-05/18/2011), 2 years and 3 months (07/01/2010-09/26/2012), and 11 years and 5 months (07/01/2010-11/23/2021). Similar to the Greenland data, the difference between the 7-meter resolution "original" and "noisy" Line of Sight (LOS) deformation is substantial and biased due to poor unwrapping. Therefore, we evaluate the performance of the LOS error using 56-meter resolution data. Figure \ref{err_snr_hist6_sanand1_fig}(a) shows the 2D histogram of the LOS deformation difference between the "original" and "noisy" data versus SNR for $0.6 <= \rho <0.7$. The data were acquired over the San Andreas Fault between 07/01/2010 and 09/28/2010. The red curve represents the exponential fit to the 2D LOS difference. As expected, the LOS error decreases as backscattered power increases. Figure \ref{err_snr_hist6_sanand1_fig}(b) shows the fitting curves for LOS deformation error at different correlation levels, including the fit curve from Figure \ref{err_snr_hist6_sanand1_fig}(a). As seen in this figure, as interferometric correlation decreases, higher backscattered power (more S/N) is required to achieve the same LOS error. Theoretically, the standard deviation of the LOS displacement error can be estimated from the interferometric correlation. Assuming an "original" total correlation, we compute the "noisy" total correlation as follows:
\begin{equation}\label{gamma_tot_eq}
\rho^{\text{noisy}}_{\text{total}} = \rho^{\text{original}}_{\text{total}} \times \rho^{\text{noisy}}_{\text{SNR}}
\end{equation}
where $\rho^{\text{noisy}}_{\text{total}}$ and $\rho^{\text{original}}_{\text{total}}$ are the total correlations of the noisy and original signals, respectively \cite{zebker1992decorrelation}. The term $\rho^{\text{noisy}}_{\text{SNR}}$ is the correlation due to the signal-to-noise ratio (SNR) and is given by:
\begin{equation}\label{rho_snr_eq}
\rho^{\text{noisy}}_{\text{SNR}} = \frac{1}{1+\frac{1}{\text{SNR}}}
\end{equation}

The standard deviation of the LOS error due to noise is then computed using the following expression:
\begin{equation}\label{std_los_err_eq}
\sigma^{\text{noisy}}_{\text{LOS}} = \frac{\lambda}{4\pi} \cdot \frac{1}{\sqrt{2N_{\text{looks}}}} \sqrt{\frac{1 - \left( \rho^{\text{noisy}}_{\text{total}} \right)^2}{\left( \rho^{\text{noisy}}_{\text{total}} \right)^2}}
\end{equation}
where $\sigma^{\text{noisy}}_{\text{LOS}}$ is the LOS displacement uncertainty, $\lambda$ is the radar signal wavelength, and $N_{\text{looks}}$ is the equivalent number of looks. This relationship is derived from the Cramér–Rao bound and is widely used in InSAR error analysis \cite{rosen2000}. For UAVSAR single-look complex images used in this study, the number of looks is $96 \times 24$. For any given $\rho^{total}_{original}$ and "noisy" SNR, we can calculate $\sigma^{noisy}_{los}$ using equations \ref{gamma_tot_eq}, \ref{rho_snr_eq}, and \ref{std_los_err_eq}. Figure \ref{err_snr_hist6_sanand1_fig} (c) shows $\sigma^{noisy}_{los}$ vs SNR for different $\rho^{original}_{total}$. As seen in this figure, the theoretical LOS standard deviation error is slightly larger than the fit shown in figure \ref{err_snr_hist6_sanand1_fig}(b), but comparable.
\begin{figure*}[!hbp]
\begin{center}
\includegraphics[width=\linewidth]{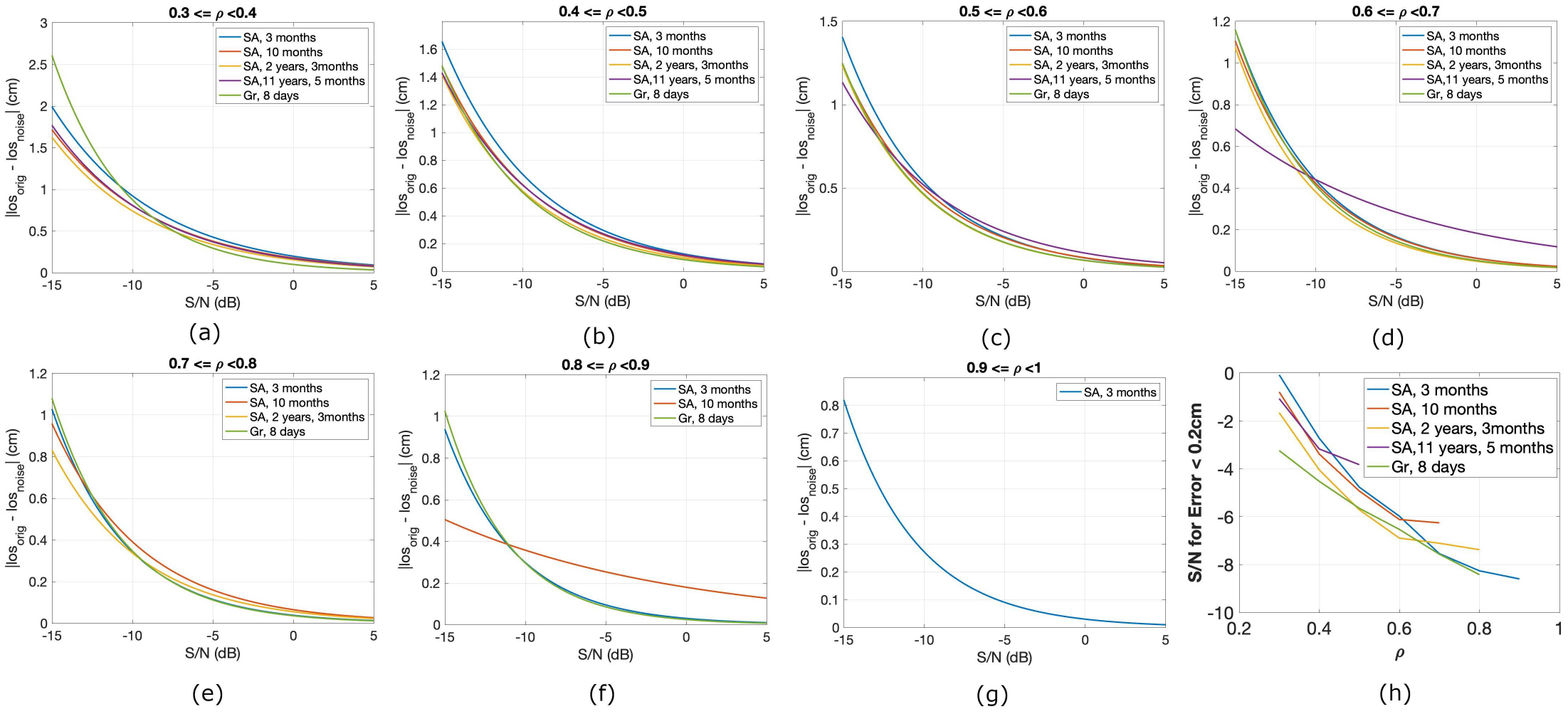}
\caption{Exponential fit to the 2D LOS difference between original and noisy deformation versus SNR for (a) $0.3 <= \rho <0.4$ (b) $0.4 <= \rho <0.5$ (c) $0.5 <= \rho <0.6$ (d) $0.6 <= \rho <0.7$ (e) $0.7 <= \rho <0.8$ (f) $0.8 <= \rho <0.9$ (g) $0.9 <= \rho < 1.0$. (h) SNR threshold for LOS difference between original and low NESZ data less than 2cm versus correlation. Blue, red, orange, and purple curves in figure a)-(h) are for San Andreas fault data between 07/01/2010-09/28/2010, 07/01/2010-05/18/2011, 07/01/2010-09/26/2012, and 07/01/2010-11/23/2021, respectively. The green curve is for Greenland data between 05/21/2009-05/29/2009.} \label{err_snr_cor_all_fig}
\end{center}
\end{figure*}

Figures \ref{err_snr_cor_all_fig} (a)--(g)  show the exponential fitting curves for the 2D histogram of the difference between the “noisy” and “original” Line of Sight (LOS) for different interferometric correlation intervals ranging from 0.3--0.4 to 0.9--1, respectively. Different colors represent different single interferograms. As mentioned earlier, we used Greenland data with 8-days separation, along with four different time intervals in the San Andreas Fault time series data. In Figures \ref{err_snr_cor_all_fig} (f) and (g), which correspond to higher correlation, we lack enough samples from certain single interferograms, such as the 2-year interferogram over the San Andreas Fault. As observed in these plots, the fitted curves for different single interferograms exhibit relatively similar behavior. The variability in LOS error due to correlation is greater than the variability due to location. Note that all results in Figure \ref{err_snr_cor_all_fig} are for multilooked, 56m resolution data. The number of looks in each direction is 48 looks for a squared pixel. Therefore, for SDC mission single look image with 5m resolution, the 48 looks results in 240m. The resulted resolution may not be sufficient to capture all the details of deformation for many science or applications.

Based on the fitted curve for the 3-month San Andreas Fault interferogram in Figure \ref{err_snr_cor_all_fig}(g), an SNR greater than -8.6 dB is required to achieve a 2 mm or better LOS error for correlation greater than 0.9. The 2 mm accuracy threshold is selected based on NASA-ISRO SAR Mission (NISAR) Co-Seismic deformation accuracy requirement \cite{nisar_handbook}. The NISAR requirements states 4mm accuracy for Co-Seismic requirement in 50km distance. If we assume half of this error is due to low NES0, 2mm difference between "original" and "noisy" is required.
We use this metric to quantify the required SNR for achieving a LOS error of less than 2 mm across the different curves in Figures 5(a)–(g). The blue curve in Figure \ref{err_snr_cor_all_fig}(h) illustrates the required SNR for a 3-month separation San Andreas Fault interferogram, plotted against correlation, to achieve a LOS error of less than 2 mm. Different colors represent various single interferograms from this study. As seen in this figure, the different interferograms show relatively similar patterns. As correlation decreases, a higher SNR is required to maintain a 2 mm LOS accuracy. However, for coherence values greater than 0.6, an SNR above -6 dB is sufficient to keep the LOS error under 2 mm for different interferograms. The noise level of our system is -15 dB. Therefore, the minimum required backscattered power for 2 mm LOS accuracy, with coherence greater than 0.6, is -21 dB. Note that our quantification is based on the fitting curve to the error. Additionally, we used 8x8 looks to achieve this accuracy. Achieving better accuracy would require more restricted backscattered power.

\subsection{Performance of Interferogram Stacks for a low NESZ Radar} \label{sec-results_stack_int}
\begin{figure}[!hbp]
\centering
\includegraphics[width=\linewidth]{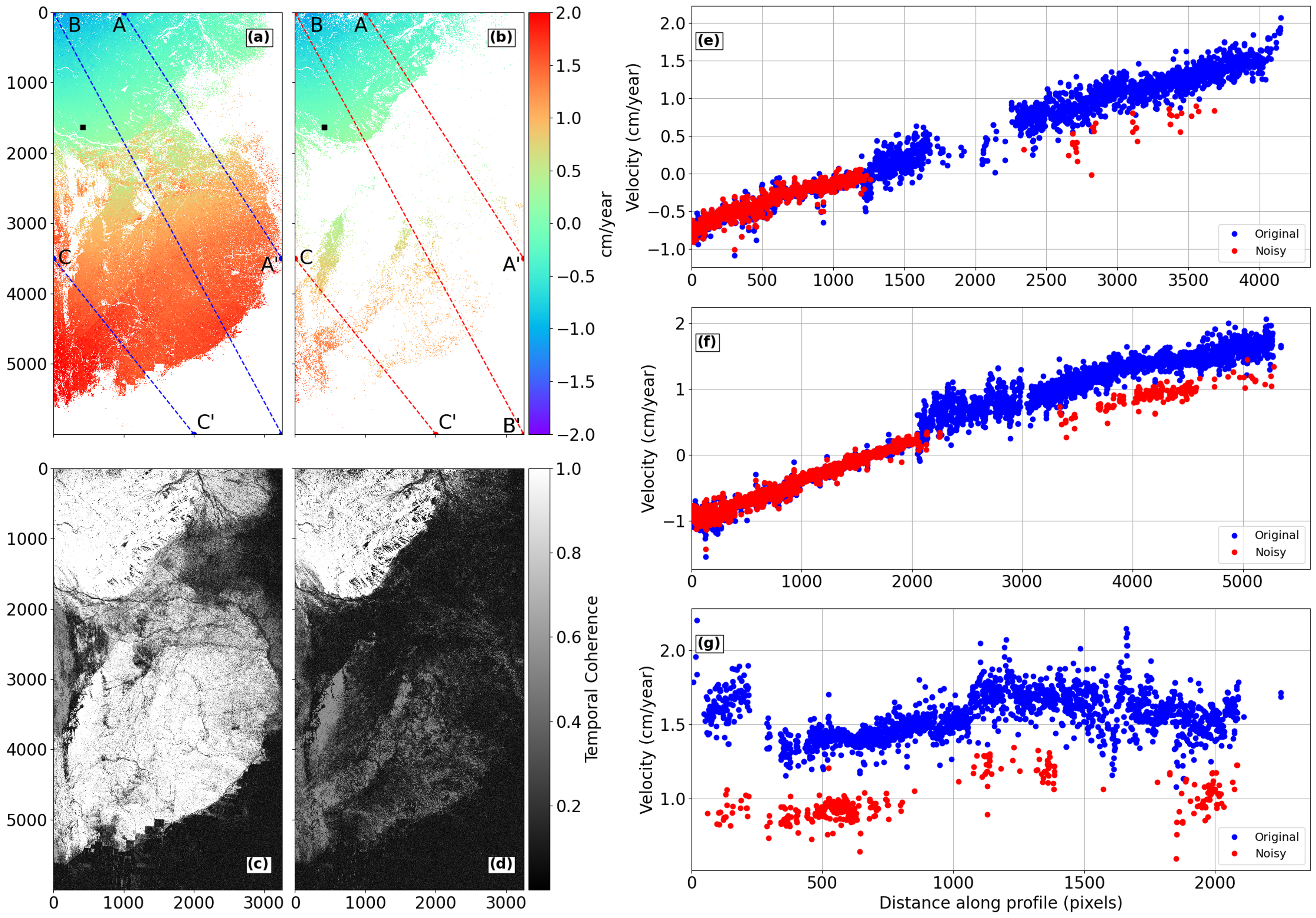}
\caption{(a) and (b) show the temporal coherence of the ``original'' and ``noisy'' stacks, respectively. (c) and (d) are the ``original'' and ``noisy'' velocity maps, respectively. The velocity fields are masked with temporal coherence threshold of 0.5 from their corresponding temporal coherence maps (a and b). The black square indicates the reference pixel and the dashed lines indicate the locations of velocity profiles. (e), (f), and (g) are velocity profiles extracted from the masked velocity fields, where blue dots represent the ``original'' stack and red dots the ``noisy'' stack at corresponding locations.
}
\label{fig-vels}
\end{figure}

Time series inversion enables the estimation of deformation history for each pixel at every acquisition relative to a reference date. We selected the earliest acquisition as the temporal reference and computed the time series for each pixel. Pixel velocities were derived by calculating the best-fitting line through the time series (Fig.~\ref{fig-vels}).

To evaluate the reliability of the time series, we calculated temporal coherence following Pepe and Lanari (2006) \cite{pepe2006extension}. Temporal coherence serves as a quality metric for the inversion process, derived by comparing the original interferograms with the reconstructed ones post-inversion. This normalized difference acts as a proxy for phase unwrapping quality at each pixel. High temporal coherence indicates minimal phase noise and higher displacement accuracy, while low coherence suggests unreliable unwrapping, potentially distorting long-term deformation trends. To ensure robust results, we applied a temporal coherence threshold of 0.5, masking low-quality pixels and excluding them from further analysis.

Figures \ref{fig-vels} (a) and (b) show the temporal coherence metric maps for the original and noisy time series, respectively. As observed, the noisy time series exhibits a significant loss of temporal coherence across many regions. Figures \ref{fig-vels} (c) and (d) present the velocity estimates for the original and noisy time series, masked using a temporal coherence threshold of 0.5.

To evaluate differences between the original and noisy velocity fields, we extracted three profiles across the velocity maps. The A–A' profile, shown in Figure \ref{fig-vels} (e), starts near the reference pixel and demonstrates strong agreement between the two velocity estimates until encountering a low-coherence region, where the noisy stack shows fewer valid pixels and increased deviations. Similarly, the B–B' profile in Figure \ref{fig-vels} (f), which crosses the scene diagonally, exhibits good agreement near the reference pixel but increasing discrepancies toward B'. Both profiles also reveal a pronounced linear ramp in the estimated velocities.

Although ramp removal is typically straightforward in InSAR studies, it involves modeling and removing a linear surface, which could bias our subsequent error analysis. Therefore, we chose not to remove the ramps, as our analysis focuses on the differences between the velocity fields, where the presence of a shared ramp minimally impacts the results.

Finally, the C–C' profile in Figure \ref{fig-vels} (g), located farther from the reference pixel, shows a consistent velocity offset of approximately $0.5~\mathrm{cm/yr}$ between the original and noisy velocity fields. Given that no differences exist in the processing methods or corrections applied to either SLC stack, we attribute this shift to a bias introduced into the noisy SLC stack.

\begin{figure}
\centering
\includegraphics[width=\linewidth]{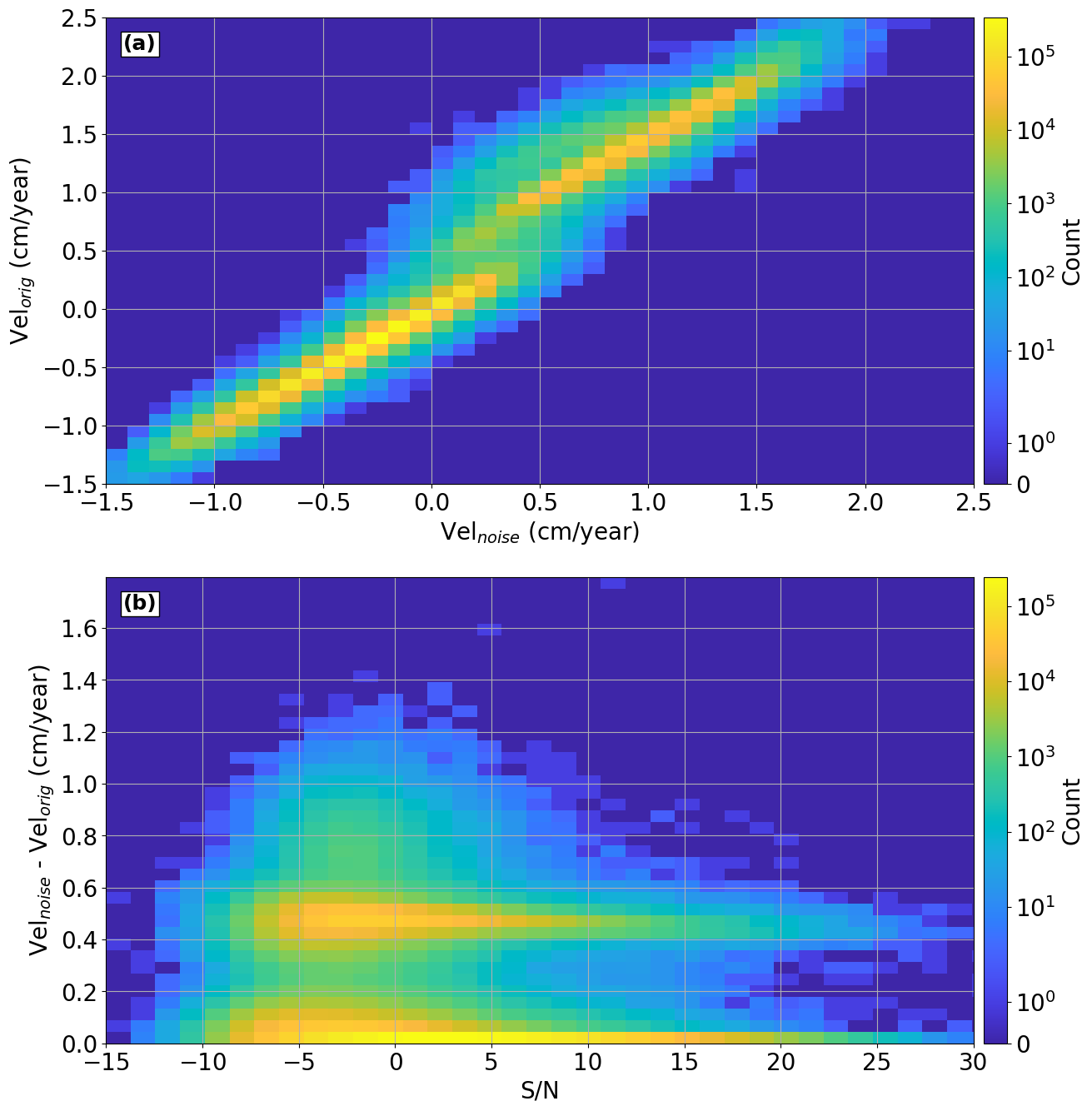}
\caption{(a) 2D histogram comparing velocities from the ``original'' and ``noisy'' stacks, highlighting the observed $0.5~\mathrm{cm/yr}$ shift. (b) 2D histogram of velocity differences vs. SNR, showing clusters around $0~\mathrm{cm/yr}$ and $0.5~\mathrm{cm/yr}$, concentrated near $-12~\mathrm{dB}$.}
\label{fig-vel_snr}
\end{figure}

Further analysis of the velocity fields through 2D histograms (Fig.~\ref{fig-vel_snr}a) confirms the consistent $0.5~\mathrm{cm/yr}$ shift observed in the profiles for original velocities greater than $0.5~\mathrm{cm/yr}$. When comparing the velocity differences between the ``original'' and ``noisy'' stacks against SNR (Fig.~\ref{fig-vel_snr}b), we observe clustering around $0~\mathrm{cm/yr}$ and $0.5~\mathrm{cm/yr}$. The discontinuity between the top and middle part of the "noisy" temporal coherence in figure \ref{fig-vels} (d), causes the phase ambiguity error of $0.5~\mathrm{cm/yr}$.

\subsubsection{Multilooking}
The $0.5~\mathrm{cm/yr}$ error in the estimated velocity can be resolved by multilooking, a standard SAR and InSAR processing technique used to improve interferogram quality. Multilooking involves averaging neighboring pixels in both the range (across-track) and azimuth (along-track) directions, which reduces speckle noise and enhances interferometric correlation. Higher coherence leads to more reliable phase measurements, improving displacement estimates. Although multilooking reduces spatial resolution due to pixel averaging, this trade-off often yields more precise velocity estimations, particularly under low-SNR conditions like those observed in our study.

\begin{figure}[!hbp]
\centering
\includegraphics[width=\linewidth]{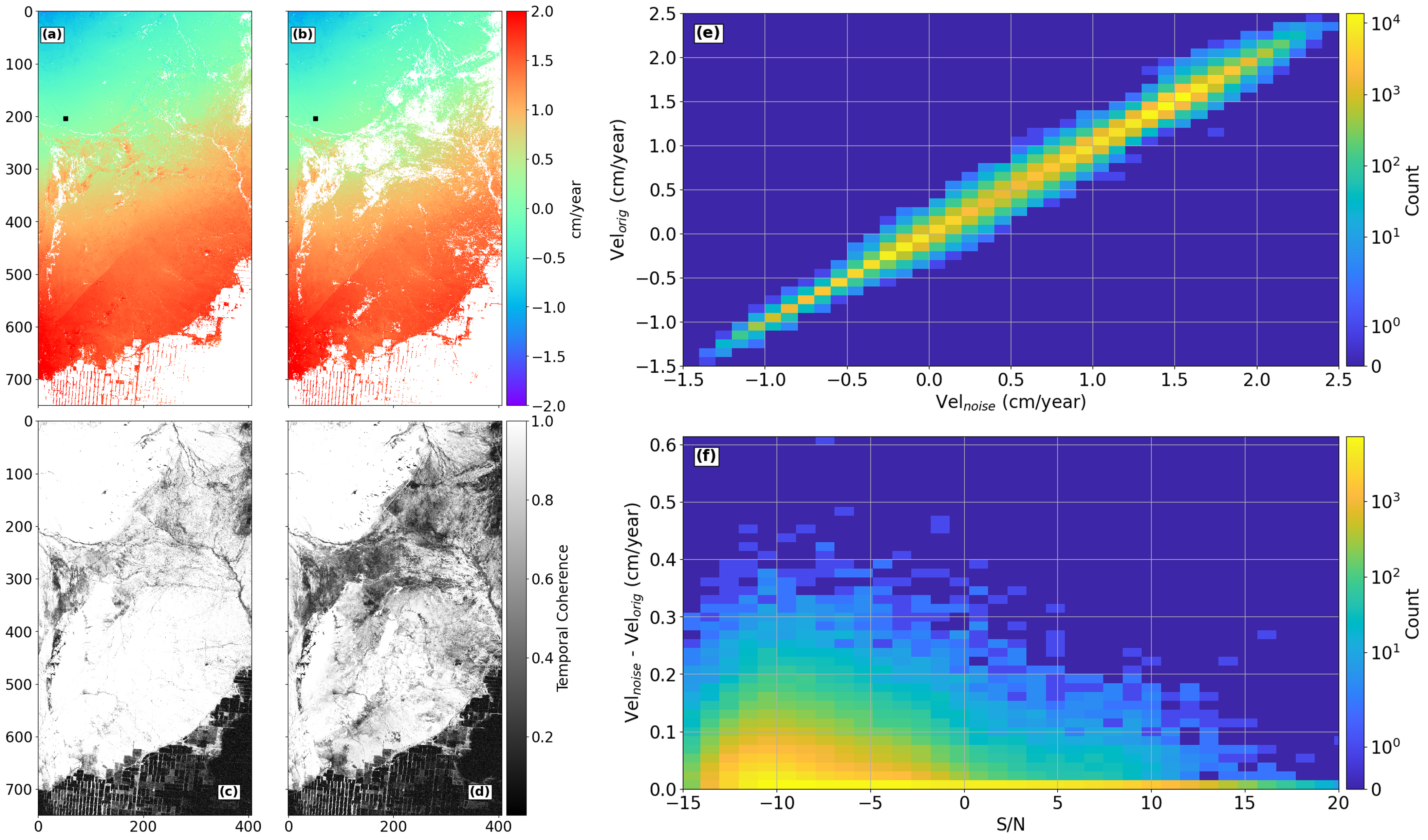}
\caption{(a) and (b) display the velocity fields after multilooking and masking with a temporal coherence threshold of 0.5. (c) and (d) show the estimated temporal coherence after multilooking for the original and noisy data stacks, respectively. (e) is a 2D histogram comparing both velocity fields, and (f) is a 2D histogram showing velocity differences relative to SNR.
}
\label{fig-vel_ml_snr}
\end{figure}

To evaluate the effectiveness of multilooking, we applied an 8x8 window to both the orginal and noisy stacks, reducing the resolution to approximately $56~\mathrm{m}$, and reprocessed the data following the workflow outlined in Section~\ref{sec-method}.
Figures \ref{fig-vel_ml_snr} (a) and (b) show the multilooked temporal coherence of "orginal" and "noisy" data, respectively.
As seen in figure Fig.\ref{fig-vel_ml_snr} (b), the multilooking significantly improved "noisy" temporal coherence as expected. Figures \ref{fig-vel_ml_snr} (c) and (d) show the multilooked estimated "original" and "noisy" velocity, respectively. The $0.5~\mathrm{cm/yr}$ velocity shift observed in the non-multilooked data was eliminated after multilooking as seen in Figure \ref{fig-vel_ml_snr}(e).

Figure \ref{fig-vel_ml_snr} (f) shows the difference between "original" and "noisy" velocities vs. SNR. While the velocity differences remain clustered around $-12~\mathrm{dB}$, the absence of systematic bias in the multilooked data indicates that multilooking not only improves coherence but also stabilizes phase measurements. This improvement highlights the trade-off between spatial resolution and measurement accuracy, demonstrating that multilooking is a practical approach for enhancing InSAR performance in low-SNR environments. The red line in figure \ref{fig-vel_ml_snr}(f) shows the exponential fit to the velocity error vs. SNR.

\begin{figure*}[!hbp]
\centering
\includegraphics[width=\linewidth]{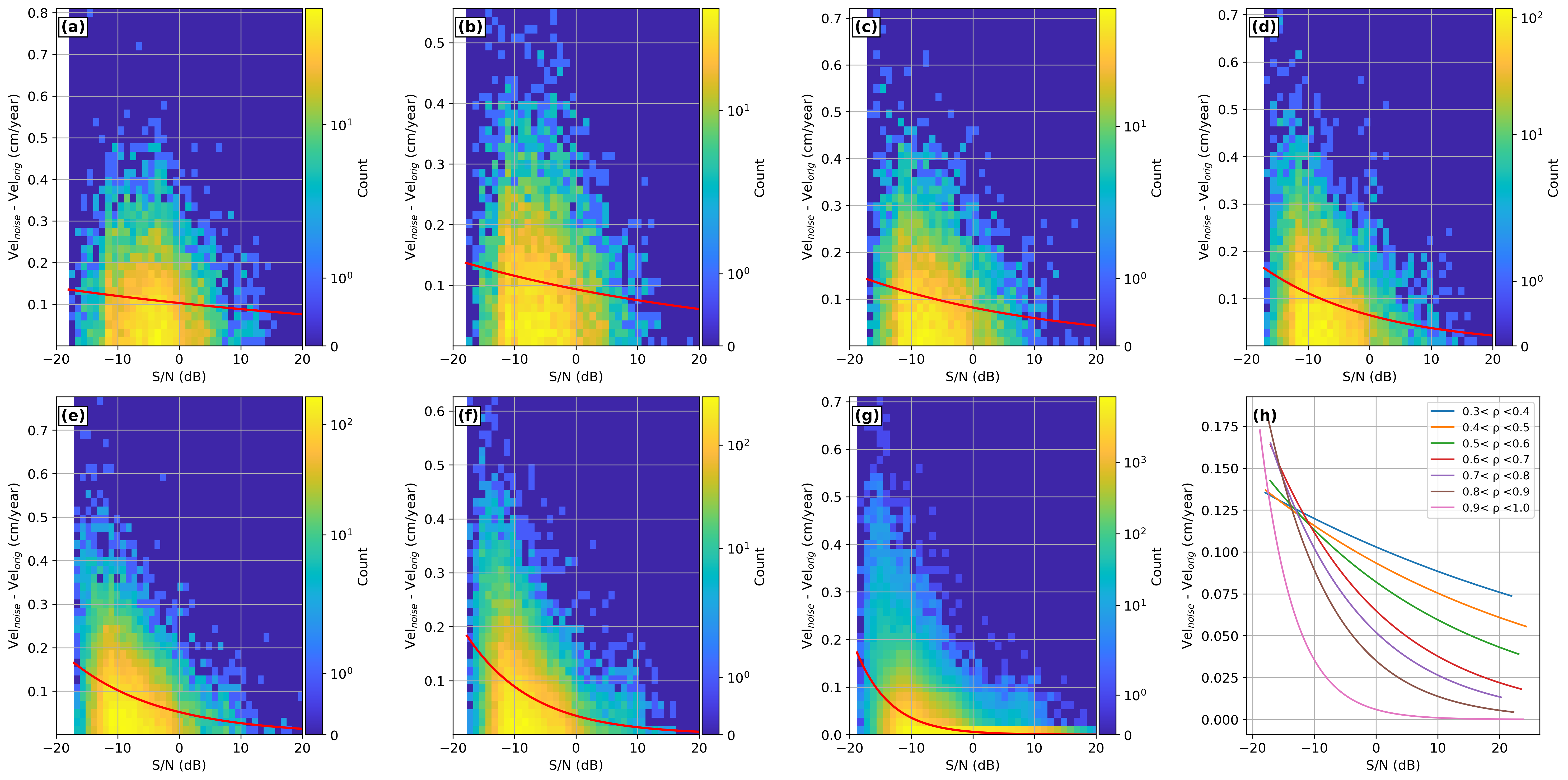}
\caption{Plots a-g show the distribution of velocity differences against SNR values for different temporal coherence ranges starting from 0.3 and reaching 1 with steps of 0.1. The red lines show the best fitting exponential function. Plot (h) shows all the exponential fits for each temporal coherence range together.}
\label{fig-vel_snr_summary}
\end{figure*}

We also investigated the relationship between velocity differences and SNR for different ranges of time series temporal coherence (Fig. \ref{fig-vel_snr_summary}). Fig. \ref{fig-vel_snr_summary}(a–g) show the distribution of velocity differences as a function of SNR, grouped into temporal coherence bins of width 0.1 starting from 0.3. These results are strongly influenced by the number of valid pixels available within each temporal coherence range.

As observed in the distributions, the majority of pixels exhibit temporal coherence values between 0.9 and 1.0. Fig. \ref{fig-vel_snr_summary}(h) summarizes the exponential fits corresponding to each coherence range. As temporal coherence increases, the velocity differences between the original and noisy time series decrease for a given SNR, as expected.


The NISAR requirement for secular deformation accuracy is $0.2~\mathrm{cm/yr}$ or better for distances less than 50km \cite{nisar_handbook}. We analyzed a $42~\mathrm{km}$ segment of a UAVSAR flight line over the San Andreas Fault Zone, selecting a coherent reference point consistent between original and noisy datasets. The farthest evaluated pixel was approximately $36.5~\mathrm{km}$ from this reference point. In order to achieve $0.2~\mathrm{cm/yr}$ accuracy, the added noise due to low-SNR should be considered along with error sources such as troposphere and ionosphere. Even by allocating half of the entire error to be due to low SNR, the SNR should be between $-17$ and $1~\mathrm{dB}$, depending on the temporal coherence (Fig.\ref{fig-vel_snr_summary}(h)). It is important to note that the fitting is less reliable for SNR values below $-12~\mathrm{dB}$ due to the limited number of available points.


These findings suggest that low-SNR SAR systems can achieve velocity precision comparable to high-SNR systems when multilooking is applied, albeit at the cost of reduced spatial resolution. The results of our study has important implications for designing future cost-effective SAR missions such as SDC, where balancing resolution and accuracy is critical.

\section{Conclusion}

This study evaluates the impact of low Signal-to-Noise Ratio (SNR) on the precision of InSAR-derived displacement measurements using UAVSAR data. Our results were benchmarked against the displacement measurement requirements of the NASA-ISRO SAR Mission (NISAR) \cite{nisar_handbook}. Key findings are as follows:

\begin{enumerate}
    \item Our analysis demonstrates that an interferometric SAR system needs SNR greater than $-6~\mathrm{dB}$ to achieve a displacement error less than $2~\mathrm{mm}$ in individual interferograms for correlation greater than 0.6, satisfying NISAR's Solid Earth Science Level-2 coseismic displacement requirement of $4~\mathrm{mm}$ at 50 km \cite{nisar_handbook}.

    \item Time-series analysis under the same low-SNR conditions indicates that the error contribution to secular deformation estimations is approximately $0.1~\mathrm{cm/yr}$ or less for SNR greater than 1dB, meeting NISAR's requirement of $0.2~\mathrm{cm/yr}$ or better at 50 km \cite{nisar_handbook}. This result indicates that reliable deformation rate estimations remain feasible under significantly degraded signal conditions.

    \item Applying multilooking with an 8$\times$8 window substantially improved interferometric coherence and effectively eliminated the systematic velocity bias of approximately $0.5~\mathrm{cm/yr}$ observed in original data. This result highlights multilooking as an effective technique for mitigating phase noise and enhancing displacement accuracy in low-SNR scenarios.

    \item While multilooking improves measurement precision, it reduces the resolution to approximately $56~\mathrm{m}$. The $56~\mathrm{m}$ resolution for multilooked UAVSAR data is equivalent with $240~\mathrm{m}$ resolution for the future SDC mission with single look image resolution of $5~\mathrm{m}$. This trade-off between precision and resolution must be carefully considered in SAR system design and data processing strategies.

    \item Results obtained from both the San Andreas Fault and Greenland ice sheet datasets exhibited consistent trends, underscoring the robustness of our findings across diverse geophysical environments. This consistency suggests broader applicability of these results for InSAR monitoring of both tectonic and glacial deformation.

    \item Achieving high-precision measurements with low-SNR SAR data indicates that future missions can optimize sensor design for cost-effectiveness. By carefully balancing system sensitivity with spatial resolution, initiatives like NASA's Surface Deformation and Change (SDC) mission can deliver high-quality deformation measurements while minimizing overall mission costs.

\end{enumerate}

\section*{Acknowledgments}
This project is part of the Surface Deformation and Change (SDC) Mission Study led by the Jet Propulsion Laboratory, California Institute of Technology, under a contract with the National Aeronautics and Space Administration (80NM0018D0004). The following NASA Centers collaborated and contributed to the project’s findings: Ames Research Center, Goddard Spaceflight Center, Langley Research Center, and Marshall Spaceflight Center.

\bibliographystyle{IEEEtran}
\bibliography{references}

@article{massonnet1998radar,
  title={Radar interferometry and its application to changes in the Earth's surface},
  author={Massonnet, Didier and Feigl, Kurt L},
  journal={Reviews of geophysics},
  volume={36},
  number={4},
  pages={441--500},
  year={1998},
  publisher={Wiley Online Library}
}

@book{hanssen2001radar,
  title={Radar interferometry: data interpretation and error analysis},
  author={Hanssen, Ramon F},
  volume={2},
  year={2001},
  publisher={Springer Science \& Business Media}
}

@article{zebker1992decorrelation,
  title={Decorrelation in interferometric radar echoes},
  author={Zebker, Howard A and Villasenor, John and others},
  journal={IEEE Transactions on geoscience and remote sensing},
  volume={30},
  number={5},
  pages={950--959},
  year={1992},
  publisher={Citeseer}
}

@article{goldstein1988satellite,
  title={Satellite radar interferometry: Two-dimensional phase unwrapping},
  author={Goldstein, Richard M and Zebker, Howard A and Werner, Charles L},
  journal={Radio science},
  volume={23},
  number={4},
  pages={713--720},
  year={1988},
  publisher={AGU}
}

@article{rocca2007modeling,
  title={Modeling interferogram stacks},
  author={Rocca, Fabio},
  journal={IEEE Transactions on Geoscience and Remote Sensing},
  volume={45},
  number={10},
  pages={3289--3299},
  year={2007},
  publisher={IEEE}
}

@article{lu2007insar,
  title={InSAR imaging of volcanic deformation over cloud-prone areas--Aleutian Islands},
  author={Lu, Zhong},
  journal={Photogrammetric Engineering \& Remote Sensing},
  volume={73},
  number={3},
  pages={245--257},
  year={2007},
  publisher={American Society for Photogrammetry and Remote Sensing}
}

@article{fore2015uavsar,
  title={UAVSAR polarimetric calibration},
  author={Fore, Alexander G and Chapman, Bruce D and Hawkins, Brian P and Hensley, Scott and Jones, Cathleen E and Michel, Thierry R and Muellerschoen, Ronald J},
  journal={IEEE Transactions on Geoscience and Remote Sensing},
  volume={53},
  number={6},
  pages={3481--3491},
  year={2015},
  publisher={IEEE}
}

@inproceedings{hawkins2018application,
  title={Application of UAVSAR data to NISAR calibration and validation},
  author={Hawkins, Brian and Riel, Bryan and Zheng, Yang and Hensley, Scott},
  booktitle={CEOS Cal/Val SAR subgroup workshop},
  year={2018}
}

@article{chen2000network,
  title={Network approaches to two-dimensional phase unwrapping: intractability and two new algorithms},
  author={Chen, Curtis W and Zebker, Howard A},
  journal={JOSA A},
  volume={17},
  number={3},
  pages={401--414},
  year={2000},
  publisher={Optica Publishing Group}
}

@article{chen2001two,
  title={Two-dimensional phase unwrapping with use of statistical models for cost functions in nonlinear optimization},
  author={Chen, Curtis W and Zebker, Howard A},
  journal={JOSA A},
  volume={18},
  number={2},
  pages={338--351},
  year={2001},
  publisher={Optica Publishing Group}
}

@article{chen2002phase,
  title={Phase unwrapping for large SAR interferograms: Statistical segmentation and generalized network models},
  author={Chen, Curtis W and Zebker, Howard A},
  journal={IEEE Transactions on Geoscience and Remote Sensing},
  volume={40},
  number={8},
  pages={1709--1719},
  year={2002},
  publisher={IEEE}
}

@article{yunjun2019small,
  title={Small baseline InSAR time series analysis: Unwrapping error correction and noise reduction},
  author={Yunjun, Zhang and Fattahi, Heresh and Amelung, Falk},
  journal={Computers \& Geosciences},
  volume={133},
  pages={104331},
  year={2019},
  publisher={Elsevier}
}

@article{pepe2006extension,
  title={On the extension of the minimum cost flow algorithm for phase unwrapping of multitemporal differential SAR interferograms},
  author={Pepe, Antonio and Lanari, Riccardo},
  journal={IEEE Transactions on Geoscience and remote sensing},
  volume={44},
  number={9},
  pages={2374--2383},
  year={2006},
  publisher={IEEE}
}

@article{nisar_handbook,
  title={mission science users’ handbook},
  author={NASA-ISRO, SAR},
  journal={Jet Propulsion Lab., California Inst. Technol., Pasadena, CA, USA, https://nisar. jpl. nasa. gov/system/documents/files/26\_NISAR\_ FINAL\_9-6-19. pdf (last access: 23 September 2022)},
  year={2018}
}

@article{wright2004insar,
  title={InSAR observations of low slip rates on the major faults of western Tibet},
  author={Wright, Tim J and Parsons, Barry and England, Philip C and Fielding, Eric J},
  journal={science},
  volume={305},
  number={5681},
  pages={236--239},
  year={2004},
  publisher={American Association for the Advancement of Science}
}

@article{bekaert2016network,
  title={A network inversion filter combining GNSS and InSAR for tectonic slip modeling},
  author={Bekaert, DPS and Segall, Paul and Wright, Tim J and Hooper, Andrew J},
  journal={Journal of Geophysical Research: Solid Earth},
  volume={121},
  number={3},
  pages={2069--2086},
  year={2016},
  publisher={Wiley Online Library}
}

@article{leinss2021tandem,
  title={TanDEM-X: Deriving InSAR height changes and velocity dynamics of great aletsch glacier},
  author={Leinss, Silvan and Bernhard, Philipp},
  journal={IEEE Journal of Selected Topics in Applied Earth Observations and Remote Sensing},
  volume={14},
  pages={4798--4815},
  year={2021},
  publisher={IEEE}
}

@article{feng2023improving,
  title={Improving the capability of D-InSAR combined with offset-tracking for monitoring glacier velocity},
  author={Feng, Xiaoman and Chen, Zhuoqi and Li, Gang and Ju, Qi and Yang, Zhibing and Cheng, Xiao},
  journal={Remote Sensing of Environment},
  volume={285},
  pages={113394},
  year={2023},
  publisher={Elsevier}
}

@article{chaussard2021over,
  title={Over a century of sinking in Mexico City: No hope for significant elevation and storage capacity recovery},
  author={Chaussard, E and Havazli, E and Fattahi, H and Cabral-Cano, E and Solano-Rojas, D},
  journal={Journal of Geophysical Research: Solid Earth},
  volume={126},
  number={4},
  pages={e2020JB020648},
  year={2021},
  publisher={Wiley Online Library}
}

@article{pacheco2015application,
  title={Application of InSAR and gravimetric surveys for developing construction codes in zones of land subsidence induced by groundwater extraction: case study of Aguascalientes, Mexico},
  author={Pacheco-Mart{\'\i}nez, J and Wdowinski, S and Cabral-Cano, E and Hern{\'a}ndez-Mar{\'\i}n, M and Ortiz-Lozano, JA and Oliver-Cabrera, T and Solano-Rojas, D and Havazli, E},
  journal={Proceedings of the International Association of Hydrological Sciences},
  volume={372},
  number={372},
  pages={121--127},
  year={2015},
  publisher={Copernicus GmbH G{\"o}ttingen, Germany}
}

@article{rosen2007uavsar,
  title={UAVSAR: New NASA airborne SAR system for research},
  author={Rosen, Paul A and Hensley, Scott and Wheeler, Kevin and Sadowy, Greg and Miller, Tim and Shaffer, Scott and Muellerschoen, Ron and Jones, Cathleen and Madsen, Soren and Zebker, Howard},
  journal={IEEE Aerospace and Electronic Systems Magazine},
  volume={22},
  number={11},
  pages={21--28},
  year={2007},
  publisher={IEEE}
}

@article{bekaert2019exploiting,
  title={Exploiting UAVSAR for a comprehensive analysis of subsidence in the Sacramento Delta},
  author={Bekaert, David PS and Jones, Cathleen E and An, Karen and Huang, Mong-Han},
  journal={Remote sensing of environment},
  volume={220},
  pages={124--134},
  year={2019},
  publisher={Elsevier}
}

@inproceedings{hensley2008uavsar,
  title={The UAVSAR instrument: Description and first results},
  author={Hensley, Scott and Wheeler, Kevin and Sadowy, Greg and Jones, Cathleen and Shaffer, Scott and Zebker, Howard and Miller, Tim and Heavey, Brandon and Chuang, Ernie and Chao, Roger and others},
  booktitle={2008 IEEE Radar Conference},
  pages={1--6},
  year={2008},
  organization={IEEE}
}

@article{ryder2008spatial,
  title={Spatial variations in slip deficit on the central San Andreas Fault from InSAR},
  author={Ryder, Isabelle and B{\"u}rgmann, Roland},
  journal={Geophysical Journal International},
  volume={175},
  number={3},
  pages={837--852},
  year={2008},
  publisher={Blackwell Publishing Ltd Oxford, UK}
}

@article{lundgren2009southern,
  title={Southern San Andreas-San Jacinto fault system slip rates estimated from earthquake cycle models constrained by GPS and interferometric synthetic aperture radar observations},
  author={Lundgren, Paul and Hetland, Eric A and Liu, Zhen and Fielding, Eric J},
  journal={Journal of Geophysical Research: Solid Earth},
  volume={114},
  number={B2},
  year={2009},
  publisher={Wiley Online Library}
}

@inproceedings{aissa2024united,
  title={United States West Coast Surface Deformation with Wide-Swath L-Band ALOS-2 PALSAR-2 and NISAR},
  author={Aissa, S Belhadj and Fielding, EJ and Liu, Z and Tymofyeyeva, E and Havazli, E and Rosen, P and Simons, Mark and Lindsay, D and B{\"u}rgmann, R and Bawden, GW},
  booktitle={IGARSS 2024-2024 IEEE International Geoscience and Remote Sensing Symposium},
  pages={2434--2437},
  year={2024},
  organization={IEEE}
}

@ARTICLE{rosen2000,
  author={Rosen, P.A. and Hensley, S. and Joughin, I.R. and Li, F.K. and Madsen, S.N. and Rodriguez, E. and Goldstein, R.M.},
  journal={Proceedings of the IEEE}, 
  title={Synthetic aperture radar interferometry}, 
  year={2000},
  volume={88},
  number={3},
  pages={333-382},
  doi={10.1109/5.838084}}

\vfill

\end{document}